\documentclass[aip,jcp,reprint,superscriptaddress,floatfix,longbibliography]{revtex4-1}
\usepackage{epsfig}
\usepackage{graphicx}
\usepackage{xcolor}
\usepackage{dcolumn} 
\usepackage{bm} 
\usepackage{amssymb}
\usepackage[version=4]{mhchem}

\newcommand{\btVFill}{\vskip0pt plus 1filll}

\begin{document}

\title{Quantifying the impact of the Tamm-Dancoff approximation on the computed spectra of transition-metal systems}

\author{Muhammed A. Dada}
\author{Sarah Pak}
\author{Matthew N. Ward}
\author{Megan Simons}
\author{Daniel R. Nascimento}
\email{daniel.nascimento@memphis.edu}
\affiliation{Department of Chemistry, The University of Memphis, Memphis, TN 38152, USA}




\begin{abstract}

The Tamm–Dancoff Approximation (TDA) offers a computationally efficient alternative to full linear-response Time-Dependent Density Functional Theory (TDDFT) for calculating electronic excited states, particularly in large molecular systems. By neglecting the coupling between excitation and de-excitation channels, TDA simplifies the TDDFT response equations into a Hermitian form. This not only reduces computational cost but also eliminates numerical instabilities that can arise in the full non-Hermitian formalism. While TDA has been widely explored for valence excitations, its reliability for transition metal complexes and core-level spectroscopies remains largely untested. In this work, we address this gap by systematically comparing TDA and full TDDFT results for a series of transition metal species, focusing on absorption spectra across the UV-Vis, metal K-edges, and L-edges. Our results show that, for core-level excitations, TDA yields excitation energies and oscillator strengths nearly indistinguishable from those obtained with full TDDFT. This agreement is attributed to the negligible contribution of de-excitation amplitudes at high excitation energies, indicating that the omitted coupling terms play an insignificant role in these spectral regimes. These findings validate the accuracy and robustness of TDA in core-level spectra simulations and support its broader application in studies involving transition metal systems.
\end{abstract}

\maketitle

\section{Introduction}
\label{IntroSection}

The linear-response formulation of time-dependent density-functional theory (TDDFT) has become a highly popular way to obtain excited-state information of large molecular systems inaccessible to high-accuracy wave-function-based theories. In its standard formulation, given by Casida's equation, one obtains single-particle density-density polarization amplitudes that can be used to evaluate linear-response properties  \cite{casida1995recent}.

It has long been acknowledged that the TDDFT formalism provides significant improvements over simple configuration-interaction singles (CIS) or time-dependent Hartree--Fock (TDHF) treatments, especially in the description of low-lying excited states \cite{Chantzis2013IsShapes,Matsuzawa2001Time-dependentRegion,Hsu2001ExcitationDecapentaene,Kadantsev2006ElectronicCalculations,Gelabert2006Charge-transferDescription}, and strongly bound excitons \cite{Byun2017ExcitonsApproximation}. Nonetheless, the coupling between excitation and de-excitation channels present in TDDFT can lead to significant problems in systems that exhibit triplet instabilities \cite{Peach2011InfluenceTDDFT}. By electing to neglect these interactions --- the Tamm-Dancoff Approximation (TDA) \cite{Hirata:1999:291} --- these problems are eliminated and significant computational savings can be achieved, often with improved results \cite{Hu2014PerformanceTheory,Andreussi2015CarotenoidsApproximation,Rangel2017AnApproximation}. Since the TDA equations share similarities with the configuration interaction singles (CIS) method, its application in the calculation of excited-state transition moments \cite{nascimento2021resonant,nascimento2022computational,Biasin:2021:Revealing,Segatta2021InDye,cavaletto2021resonant,Gu2021ManipulatingCavities} and non-adiabatic couplings \cite{Ou:2015:064114,Ou:2015:7150,Zhang:2015:064109,Alguire:2015:7140,Sheng:2020:4693} can be easily justifiable. 

A critical aspect of the TDA is that the Thomas--Reiche--Kuhn (TRK) dipole sum rule \cite{thomas1925zahl,reiche1925zahl,kuhn1925gesamtstarke} is no longer expected to hold \cite{Dreuw2005Single-referenceMolecules}, oftentimes leading to inconsistent spectral intensities. Garcia-Alvarez and Gozem recently demonstrated that violation of the TRK sum rule can yield TDA oscillator strengths that are particularly susceptible to the choice of dipole gauge (length vs. velocity) \cite{Garcia-Alvarez2024AbsorptionGauge,Garcia-Alvarez2025ImplementationStudy}. These problems can be severe for electronic circular dichroism calculations \cite{Bannwarth2015ElectronicApproximation}. Studies on confined systems able to exhibit mixed excitonic-plasmonic behaviour \cite{Gruning2009Exciton-plasmonApproximation,Rocca2014AbApproximation} also highlight limitations of the TDA. 

In the context of core-level spectroscopies, the TDA is routinely employed without significant scrutiny, as it consistently yields reliable relative excitation energies and oscillator strengths. It is generally accepted that the limitations of TDA observed at low excitation energies do not necessarily extend to the high-energy regime relevant to X-ray spectroscopy. A recent study by Fransson and Pettersson explored the influence of both the TDA and the Core–Valence Separation (CVS) scheme on the calculation of X-ray absorption and emission spectra\cite{Fransson2024EvaluatingCalculations}. Their investigation focused primarily on K-edge transitions in light elements, with selected examples involving sulfur and selenium. They concluded that X-ray absorption spectra (XAS) calculations using the TDA and CVS approximations closely reproduce the results obtained from TDDFT treatments.

In the present work, we extend this analysis to a broader set of transition-metal-containing complexes and spectral regions, providing a systematic assessment of the TDA's performance across a wide excitation energy range. Our results offer quantitative evidence that the coupling betwee excitation and de-excitation included in TDDFT diminishes rapidly with increasing excitation energy, becoming effectively negligible in the soft X-ray and higher energy regimes. These findings support and further justify the widespread use of the TDA in core-level spectra calculations.

\btVFill



\begin{figure*}[htbp!]
    \centering
    \includegraphics[width=0.8\linewidth]{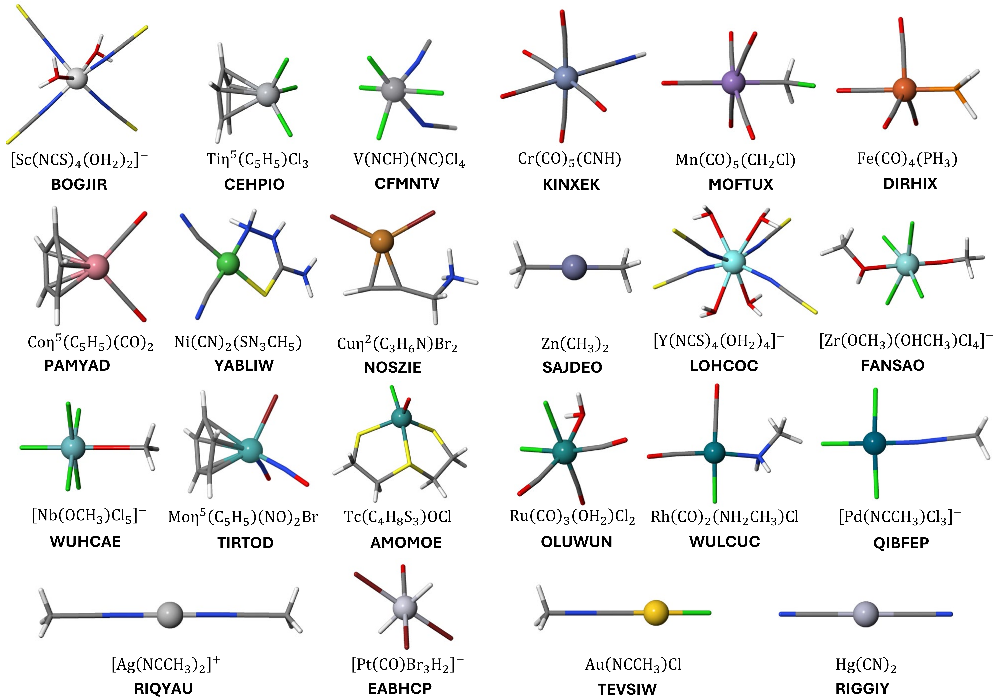}
    \caption{Collection of 22 transition metal complexes chosen for the present study. Structures were retrieved from the tmQM dataset \cite{tmQM} and are shown with their chemical formulas and Cambridge Structural Database (CSD) reference code.}
    \label{fig:geometries}
\end{figure*}
\section{Methodology}
\label{TheorySection}

The linear-response time-dependent density functional theory problem is given by Casida's equations\cite{casida1995recent}: 
\begin{equation}
    \label{EQN:Casida}
    \begin{pmatrix}
        \bf{A} & \bf{B} \\
        \bf{B} & \bf{A} 
    \end{pmatrix}
    \begin{pmatrix}
        \bf{X} \\ \bf{Y}
    \end{pmatrix}
    =
    \Omega
    \begin{pmatrix}
        \bf{1} & \bf{0}\\
        \bf{0} & -\bf{1} 
    \end{pmatrix}
    \begin{pmatrix}
        \bf{X} \\ \bf{Y} 
    \end{pmatrix},
\end{equation}
where the elements of the $\bf{A}$ and $\bf{B}$ matrices are given by
\begin{equation}
    A_{ia,jb} = (\epsilon_a-\epsilon_i)\delta_{ij}\delta_{ab} + \frac{\partial F_{ia}}{\partial \rho_{bj}} ~~~\text{and}~~~  B_{ia,jb} = \frac{\partial F_{ia}}{\partial \rho_{jb}}.
\end{equation}

Within the standard hybrid adiabatic-local-density approximation (ALDA) formalism \cite{Furche2002AdiabaticProperties}, ${\partial F_{ia}}/{\partial \rho_{jb}}$ take the explicit form
\begin{equation}
    \frac{\partial F_{ia}}{\partial \rho_{jb}} = \langle ib|aj \rangle - \alpha_{\rm HF} \langle ib|ja \rangle + f^{\rm xc}_{ia,bj},
\end{equation}
where it is understood that the exchange component of the exchange-correlation response kernel, $f^{\rm xc}$, has been appropriately scaled by $(1-\alpha_{\rm HF})$.

In solving (\ref{EQN:Casida}), one obtains a pair of solutions corresponding to excitation ($X, Y$, and $\Omega$) and de-excitation ($X^*, Y^*$, and $-\Omega$) processes, with the $X$ and $Y$ amplitudes satisfying the condition
\begin{equation}
    \label{EQN:norm}
    \langle X^{m}|X^{n}\rangle - \langle Y^{m}|Y^{n}\rangle = \pm\delta_{mn},
\end{equation}
where the positive (negative) sign holds for positive (negative)-frequency solutions.

In the Tamm-Dancoff Approximation, it is assumed that the excitation and de-excitations channels do not couple, and $\bf{B}$ is set to zero, leading to a simpler, Hermitian problem
\begin{equation}
    \bf{A} \bf{X} = \Omega \bf{X}.
\end{equation}

Hence, the magnitude of the de-excitation amplitudes vector, $\bf{Y}$, for positive-frequency solutions, is a direct reporter on the extent to which excitation and de-excitations couple via the $\bf{B}$ matrix, and thus, on how much TDA and TDDFT are expected to differ.

In this study, we evaluate the squared norm of the de-excitation amplitude vector, $|\mathbf{Y}|^2$, across a range of transition-metal systems and over a broad energy spectrum spanning the UV-Vis region as well as the metal K and L edges. We use this quantity as a metric to assess the expected validity of the Tamm–Dancoff Approximation (TDA). To quantitatively examine discrepancies between TDA and TDDFT, we also analyze several key metrics: spectral area differences, spectral energy shifts, percentage error differences, and $R^2$ correlation coefficients.


\section{Computational Details}
\label{CompSection}

All molecular geometries used in this study (Figure \ref{fig:geometries}) were retrieved from the tmQM dataset \cite{tmQM} without further optimization. Ground and excited state calculations were carried out using an in-house code interfaced to PySCF \cite{sun2018pyscf} and employed three exchange-correlation functionals: B3LYP \cite{Becke:1988:3098, Lee:1988:785}, PBE0 \cite{Perdew:1996:9982, Adamo:1999:6158}, and PBE \cite{pbe}. A mixed basis set approach was adopted, where the Sapporo-DKH-DZP-2012-diffuse \cite{Noro:2012:1124} basis set was used for all transition metal atoms, while the 6-31G* \cite{ditchfield1971a,hehre1972a,hariharan1973a} Pople basis set was applied to non-metal atoms. Scalar relativistic corrections were incorporated using the Zeroth-Order Regular Approximation (ZORA) \cite{van-Lenthe:1994:9783, Van-Wullen:1998:392, Nichols:2009:491,pak2025fast}.

Core-level absorption spectra were calculated via full diagonalization in the respective K, L$_{1}$, L$_{2,3}$, and UV-Vis subspaces as defined by the core-valance-separation approach \cite{Cederbaum:1980:206}. In the UV-Vis case, the occupied space was defined by the a set number of valence orbitals (see supplementary information). A uniform Lorentzian broadening was applied to account for core-hole lifetime effects. The full-width-at-half-maximum (FWHM) for each edge were chosen to be 1.5, 1.0, 0.5, and 0.25 eV for the K, L$_1$, L$_{23}$, and UV-Vis regions, respectively.

\section{Results and Discussion}
\label{ResultsSection}

A systematic evaluation of the accuracy of the TDA for the calculation of linear absorption spectra was conducted across a diverse range of transition metal complexes. Since an early study by some of the authors indicates that the exchange-correlation response kernel plays a negligible role in the calculation of core-level spectra \cite{pak2024role}, we chose to focus our analysis on a single, popular, functional: B3LYP. Nonetheless, our conclusions should be generalizable to any other hybrid GGA functional.

The complexes studied (shown in Figure \ref{fig:geometries}) included all first-row and most second-row transition metals (excluding cadmium due to convergence difficulties), along with prominent third-row representatives: platinum, gold, and mercury.

An initial assessment of the TDA was performed by evaluating $|\bf{Y}|^2$ for each excited state in a given spectral region. As indicated in the Theory section, the magnitude of the $\bf{Y}$ vector is a direct consequence of the coupling between excitation and de-excitation that are neglected in the TDA. These amplitudes can also be though of as introducing ground-state correlation \cite{rowe1968equations,Scuseria:2008:231101}. 
\begin{figure}
    \centering
    \includegraphics{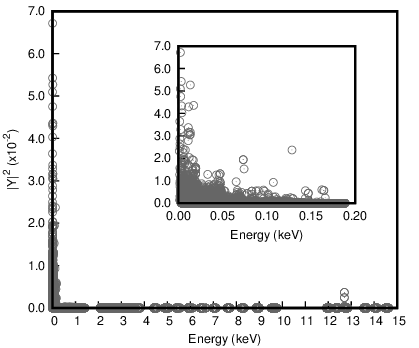}
    \caption{Magnitude square of the de-excitation amplitudes, ${\bf{Y}}^2$ as a function of energy covering the 0-15 keV range. The calculated states comprise UV-Vis, metal L, and metal K edges for all 22 complexes. The inset plot highlights the first 200 eV.}
    \label{fig:ymagnitude}
\end{figure}

Figure \ref{fig:ymagnitude} shows $|\bf{Y}|^2$ for all states calculated in a given spectral region (K, L$_1$, L$_{23}$, or UV-Vis) for all 22 complexes. As one can observe, $|\bf{Y}|^2$ is essentialy zero for all states above 200 eV and never exceeds 0.07 in the UV-Vis region. The largest concentration of states with $|\bf{Y}|^2$ above 0.02 is found at relatively low energies (Fig. \ref{fig:ymagnitude} inset), indicating that core-level excitations are in essence unaffected by the off-diagonal blocks of the TDDFT response matrix. Based on this information, one can expect the TDA and TDDFT methods to yield identical results at sufficiently high energies. Two clear outliers can, however, be observed at about 12.7 keV. The specific data points correspond to the Hg L$_{23}$ edge of \ce{Hg(CN)2}. In general, we observed that the largest values of $|\bf{Y}|^2$ belong to systems containing $\pi$-backbonding ligands. 
\begin{figure}
    \centering
    \includegraphics{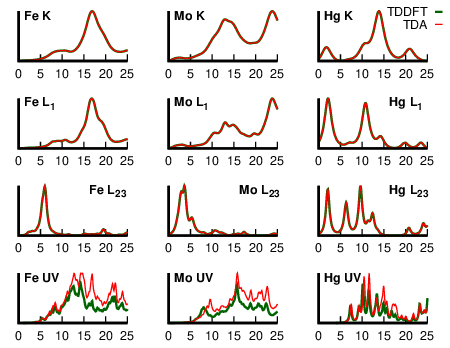}
    \caption{Calculated UV-Vis, metal K, and metal K spectra for selected complexes with and without the TDA. The plotted spectra shows the first 25 eV near the edge. X-ray plots start at 7,005 eV (Fe K), 822 eV (Fe L$_1$), 700 eV (Fe L$_{23}$), 19,715 eV (Mo K), 2,800 eV (Mo L$_1$), 2,517 eV (Mo L$_{23}$), 81,420 eV (Hg K), 14,479 eV (Hg L$_1$), and 12,692 eV (Hg L$_{23}$).  }
    \label{fig:edges}
\end{figure}

Figure \ref{fig:edges} shows the metal K, L$_1$, and L$_{23}$, and UV-Vis spectra of \ce{Fe(CO)4(PH3)}, \ce{Mo(C5H5)(NO)2Br}, and \ce{Hg(CN)2}, identified as some of the species with the largest values of $|\bf{Y}|^2$. All the core-level spectra are identical in both TDDFT and TDA cases. Appreciable differences are only visible in the UV-Vis region, where the TDA spectra shows a systematic overestimation of oscillator strengths, consistent with what has been observed in organic molecules \cite{Garcia-Alvarez2024AbsorptionGauge,Garcia-Alvarez2025ImplementationStudy}. An important corollary is that deviations from the TRK dipole sum rule resulting from the TDA should be restricted to valence-level excitations.

To better quantify any discrepancies between the two approaches, we evaluate the integrated error over a given spectral region as
\begin{equation}
\Delta I = \int\limits_{E_i}^{E_f} \frac{I_{\mathrm{TDA}}(E) - I_{\mathrm{TDDFT}}(E)}{I_{\mathrm{TDDFT}}(E)} \times 100 ~d E.
\end{equation}

\begin{figure}[!h]
    \centering
    \includegraphics[width=1\linewidth]{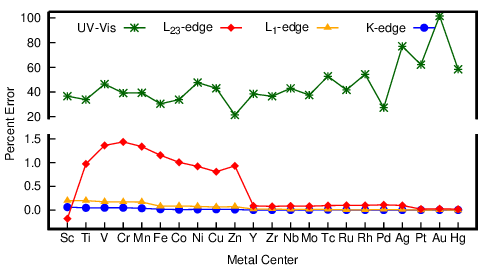}
    \caption{Percent Error of TDA relative to TDDFT for different spectral regions across the transition metal series. 
}
    \label{fig:integrated_error}
\end{figure}

This metric provides a direct assessment of the quantitative variations in both the oscillator strengths and excitation energies resulting from the two approaches. The calculated errors are summarized in Figure \ref{fig:integrated_error} and Table \ref{TAB:percent_errors}. 

\begin{table}
\centering
\setlength{\tabcolsep}{0.5em} 
{\renewcommand{\arraystretch}{1.2}
\begin{tabular}{|c |c c c c |}
\hline
\textbf{Metal} & \textbf{K} & \textbf{L$_1$} & \textbf{L$_{23}$} & \textbf{UV-Vis} \\
\hline
\hline
Sc & 0.06 & 0.20 & -0.18 & 36.54 \\
Ti & 0.05 & 0.20 &  0.97 & 33.81 \\
V & 0.05 & 0.17 & 1.36 & 46.38 \\
Cr & 0.05 & 0.17 & 1.44 & 40.54 \\
Mn & 0.04 & 0.17 & 1.34 & 39.36 \\
Fe & 0.02 & 0.08 & 1.15 & 41.82 \\
Co & 0.01 & 0.08 & 1.00 & 33.79 \\
Ni & 0.01 & 0.08 & 0.92 & 47.68 \\
Cu & 0.01 & 0.06 & 0.81 & 42.98 \\
Zn & 0.01 & 0.07 & 0.93 & 21.34 \\
\hline
\hline
Y & 0.00 & 0.02 & 0.09 & 38.47 \\
Zr & 0.00 & 0.02 & 0.08 & 36.49 \\
Nb & 0.00 & 0.01 & 0.08 & 42.90 \\
Mo & 0.00 & 0.02 & 0.09 & 39.52 \\
Tc & 0.01 & 0.01 & 0.09 & 52.74 \\
Ru & 0.00 & 0.00 & 0.10 & 41.67 \\
Rh & 0.00 & 0.00 & 0.10 & 54.36 \\
Pd & 0.00 & 0.01 & 0.11 & 27.32 \\
Ag & 0.00 & 0.01 & 0.10 & 77.00 \\
\hline
\hline
Pt & 0.00 & 0.00 & 0.02 & 62.21 \\
Au & 0.00 & 0.00 & 0.03 & 90.62 \\
Hg & 0.00 & 0.00 & 0.02 & 110.26 \\
\hline
\hline
\end{tabular}}
\caption{Percent Error differences between the TDDFT and TDA spectra for transition metal complexes evaluated at K, L$_1$, L$_{23}$, and UV-Vis spectral regions.}
\label{TAB:percent_errors}
\end{table}


At the K and L$_1$ edges, which probe excitations from the 1$s$ and 2$s$ core orbitals to unoccupied states, respectively, no significant deviations are observed. The errors are at or below 0.2\% for all systems. The first-row TMs, show the largest deviations, with percentage errors ranging from 0.06\% (Cu) to 0.2\% (Sc). Notably, the errors show a slight dependence on the atomic number, with a trend toward smaller discrepancies for heavier elements within this series, resulting from deeper core orbitals. For heavier TMs, the errors are on the order of 0.02\% or less. 

Larger discrepancies between the TDDFT and TDA spectra become more apparent at the L$_{23}$ edge, which arises from electronic transitions of metal 2$p$ core electrons to unoccupied states. Here, there is a clear distinction between the 3$d$ and 4$d$ complexes. While the 4$d$ complexes remain relatively insensitive to the TDA, the 3$d$ complexes show significantly larger errors, reaching 1.44\% for the chromium complex. Nonetheless, these discrepancies remain small enough to be negligible in any practical application.

The situation is considerably different in the UV-Vis region. In this case, the lowest error was 21\% for the zinc complex, which is sufficient to cause qualitative disagreements between the two approaches. One must note, however, that in order to allow full diagonalization, we have restricted valence excitations to a specific set of occupied orbitals. We do not expect this restriction to affect our conclusions.

\begin{figure}
    \centering
    \includegraphics{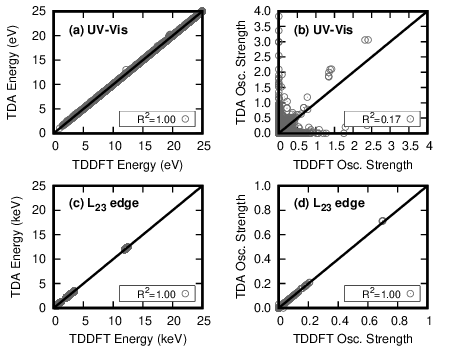}
    \caption{Correlation plots for excitation energies and oscillator strengths at different spectral regions.}
    \label{fig:correlation}
\end{figure}

To better assess whether these discrepancies are dominated primarily by errors in the excitation energies or in the oscillator strengths, we generated correlation plots at different energy ranges (Figure. \ref{fig:correlation}).

The TDA and TDDFT energies correlate well in both scenarios, with a linear correlation coefficient $R^2$ of 1. While no outliers are observed in the L$_{23}$ or deeper edges, a few outliers can be seen in the UV-Vis case, with a maximum energy deviation of 0.9 eV. Nonetheless, the average error in this region is of only 0.01 $\pm$ 0.02 eV. For oscillator strengths, large discrepancies can be observed in the UV-Vis region, with the TDA, in general, severely overestimating TDDFT. Symmetric data points about the diagonal indicate possible changes in the ordering of excited states resulting from qualitative failures of the TDA. Importantly, these problems are not observed at the L$_{23}$ and deeper edges.

\section{Summary and Conclusions}
\label{ConclusionSection}

In summary, we systematically evaluated the accuracy of the Tamm-Dancoff Approximation to TDDFT in the absorption spectra calculation of transition metal complexes across different spectral regions. Our data demonstrated that the de-excitation amplitudes from TDDFT become negligible in the high-energy ($E > 200$ eV) limit, implying that in this regime, TDDFT and TDA yield equivalent spectra. An important consequence of this behavior is that the breakdown of the Thomas--Reiche--Kuhn dipole sum rule, a major disadvantage of the TDA, is restricted to valence-level excitations. 

In the UV-Vis region, the TDA results in large errors. These errors are primarily associated with large overestimation of the oscillator strengths, as reported in earlier studies by Garcia-Alvarez and Gozem \cite{Garcia-Alvarez2024AbsorptionGauge,Garcia-Alvarez2025ImplementationStudy}. Transition metal complexes containing $\pi$-backbonding ligands are particularly affected by the couplings between excitation and de-excitation channels in TDDFT, indicating higher levels of excited-state electron correlation in these complexes.

These results further validate and support the use of the TDA in core-level spectra calculations, and reinforce that caution must be taken when employing the TDA in the calculation of UV-Vis spectra, especially if the target excited states exhibit a high-level of electron correlation.

\vspace{5pt}
{\bf Acknowledgments}

This work was supported by the National Science Foundation under the CAREER grant No. CHE-2337902 (M.~A.~D, S.~P, and D.~R.~N). M.~N.~W. and M.~S. acknowledge support by the University of Memphis. This research also benefited from computational resources provided by the University of Memphis High-Performance Computing Facility.

\vspace{5pt}
{\bf Data Availability:}
The data that support the findings of this study are available from the corresponding author upon reasonable request.

\bibliography{main,references}

\end{document}


\title{Supplementary Information: Quantifying the impact of the Tamm-Dancoff approximation on the computed spectra of transition-metal systems}

\author{Muhammed A. Dada}
\author{Sarah Pak}
\author{Matthew N. Ward}
\author{Megan Simons}
\author{Daniel R. Nascimento}
\email{daniel.nascimento@memphis.edu}
\affiliation{Department of Chemistry, The University of Memphis, Memphis, TN 38152, USA}





\maketitle





\section{Active occupied molecular orbitals defining the core-valence separation}

\begin{table}[!h]
\centering
\setlength{\tabcolsep}{0.5em} 
{\renewcommand{\arraystretch}{1.0}
\begin{tabular}{||c |c |c c c c |}
\hline
\textbf{CSD Reference Code} & \textbf{Metal} & \textbf{K} & \textbf{L$_1$} & \textbf{L$_{23}$} & \textbf{UV-Vis} \\
\hline
\hline
BOGJIR & Sc & 1 & 8 & 9--11 & 72--79 \\
CEHPIO & Ti & 1 & 5 & 6--8 & 39--54 \\
CFMNTV & V & 1 & 6 & 7--9 & 41--59 \\
KINXEK & Cr & 1 & 2 & 3--5 & 27--54 \\
MOFTUX & Mn & 1 & 3 & 4--6 & 32--60 \\
DIRHIX & Fe & 1 & 3 & 4--6 & 27--50 \\
PAMYAD & Co & 1 & 2 & 3--5 & 40--45 \\
YABLIW & Ni & 1 & 3 & 4--6 & 34--51 \\
NOSZIE & Cu & 3 & 12 & 13--15 & 53--65 \\
SAJDEO & Zn & 1 & 2 & 3--5 & 19--24 \\
\hline
\hline
LOHCOC & Y & 1 & 6 & 7--9 & 91--98 \\
FANSAO & Zr & 1 & 6 & 7--9 & 51--72 \\
WUHCAE & Nb & 1 & 7 & 8--10 & 56--72 \\
TIRTOD & Mo & 1 & 3 & 4--6 & 50--71 \\
AMOMOE & Tc & 1 & 2 & 4--6 & 53--74 \\
OLUWUN & Ru & 1 & 2 & 3--5 & 42--65 \\
WULCUC & Rh & 1 & 2 & 3--5 & 35--54 \\
QIBFEP & Pd & 1 & 2 & 3--5 & 55--60 \\
RIQYAU & Ag & 1 & 2 & 3--5 & 29--45 \\
\hline
\hline
EABHCP & Pt & 1 & 2 & 6--8 & 83--100 \\
TEVSIW & Au & 1 & 2 & 3--5 & 44--59 \\
RIGGIY & Hg & 1 & 2 & 3--5 & 39--53 \\
\hline
\end{tabular}}
\caption{the range of orbitals chosen for the calculation.}
\end{table}

\section{Error metrics for the PBE and PBE0 functionals}

\begin{figure}[h!]
    \centering
    \includegraphics[width=1\linewidth]{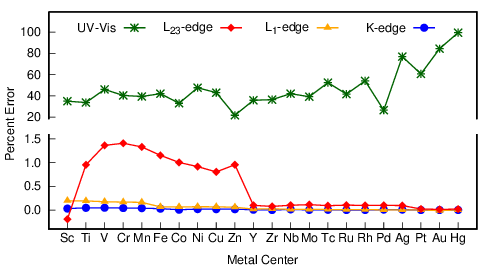}
    \caption{Percent Error of TDA relative to TDDFT for different spectral regions across the transition metal series obtained with the PBE0 functional.}
\end{figure}

\begin{table}[h!]
\centering
\setlength{\tabcolsep}{0.5em} 
{\renewcommand{\arraystretch}{1.2}
\begin{tabular}{|c|c c c c|}
\hline
\textbf{Metal} & \textbf{K} & \textbf{L$_1$} & \textbf{L$_{23}$} & \textbf{UV-Vis} \\
\hline
\hline
Sc & 0.03 & 0.19 & -0.19 & 34.86 \\
Ti & 0.05 & 0.19 & 0.95 & 33.63 \\
V & 0.05 & 0.18 & 1.36 & 45.95 \\
Cr & 0.04 & 0.17 & 1.41 & 40.28 \\
Mn & 0.04 & 0.16 & 1.33 & 39.25 \\
Fe & 0.03 & 0.07 & 1.15 & 41.95 \\
Co & 0.01 & 0.07 & 1.00 & 32.94 \\
Ni & 0.02 & 0.07 & 0.92 & 47.49 \\
Cu & 0.02 & 0.07 & 0.80 & 42.90 \\
Zn & 0.02 & 0.06 & 0.96 & 21.62 \\
\hline
\hline
Y & 0.01 & 0.02 & 0.10 & 35.78 \\
Zr & 0.00 & 0.02 & 0.08 & 36.32 \\
Nb & 0.01 & 0.01 & 0.10 & 41.93 \\
Mo & 0.00 & 0.02 & 0.11 & 39.11 \\
Tc & 0.00 & 0.01 & 0.09 & 52.31 \\
Ru & 0.00 & 0.01 & 0.10 & 41.49 \\
Rh & 0.00 & 0.01 & 0.10 & 53.96 \\
Pd & 0.01 & 0.00 & 0.10 & 26.47 \\
Ag & 0.01 & 0.00 & 0.09 & 76.85 \\
\hline
\hline
Pt & 0.00 & 0.00 & 0.02 & 60.52 \\
Au & 0.00 & -0.01 & 0.01 & 84.26 \\
Hg & 0.00 & 0.00 & 0.02 & 99.40 \\
\hline
\end{tabular}}
\caption{Percent error differences between the TDDFT and TDA spectra for transition-metal complexes evaluated at K, L$_1$, L$_{23}$, and UV-Vis spectral regions obtained with the PBE0 functional.}
\end{table}

\begin{figure}[h!]
    \centering
    \includegraphics[width=1\linewidth]{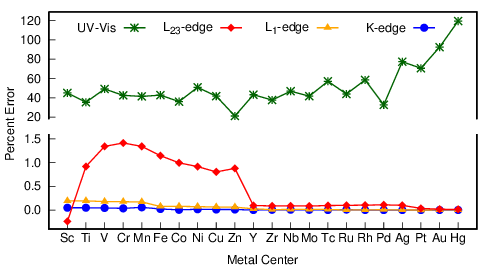}
    \caption{Percent Error of TDA relative to TDDFT for different spectral regions across the transition metal series obtained with the PBE functional.}
\end{figure}

\begin{table}[h!]
\centering
\setlength{\tabcolsep}{0.5em} 
{\renewcommand{\arraystretch}{1.2}
\begin{tabular}{|c|c c c c|}
\hline
\textbf{Metal} & \textbf{K} & \textbf{L$_1$} & \textbf{L$_{23}$} & \textbf{UV-Vis} \\
\hline
\hline
Sc & 0.05 & 0.19 & -0.24 & 45.08 \\
Ti & 0.05 & 0.19 & 0.92 & 35.45 \\
V & 0.04 & 0.18 & 1.34 & 49.14 \\
Cr & 0.03 & 0.18 & 1.41 & 42.59 \\
Mn & 0.06 & 0.17 & 1.34 & 41.49 \\
Fe & 0.02 & 0.07 & 1.15 & 42.80 \\
Co & 0.00 & 0.08 & 0.99 & 36.12 \\
Ni & 0.02 & 0.07 & 0.92 & 50.83 \\
Cu & 0.01 & 0.06 & 0.80 & 41.73 \\
Zn & 0.01 & 0.06 & 0.88 & 21.21 \\
\hline
\hline
Y & 0.00 & 0.03 & 0.09 & 43.24 \\
Zr & 0.00 & 0.01 & 0.09 & 37.74 \\
Nb & 0.01 & 0.01 & 0.09 & 46.88 \\
Mo & 0.00 & 0.01 & 0.08 & 41.68 \\
Tc & 0.00 & 0.01 & 0.09 & 57.08 \\
Ru & 0.01 & 0.00 & 0.10 & 43.98 \\
Rh & 0.00 & 0.00 & 0.10 & 58.54 \\
Pd & 0.00 & 0.01 & 0.11 & 32.72 \\
Ag & 0.00 & 0.00 & 0.10 & 77.42 \\
\hline
\hline
Pt & 0.00 & 0.00 & 0.03 & 70.63 \\
Au & 0.00 & 0.00 & 0.02 & 92.52 \\
Hg & 0.00 & 0.00 & 0.01 & 119.40 \\
\hline
\end{tabular}}
\caption{Percent error differences between the TDDFT and TDA spectra for transition-metal complexes evaluated at K, L$_1$, L$_{23}$, and UV-Vis spectral regions obtained with the PBE functional.}
\end{table}